# Hypothetical Gravity Control and Possible Influence on Space Propulsion

Tajmar, M.[*]

*Space Propulsion, ARC Seibersdorf research, A-2444 Seibersdorf, Austria*

O. Bertolami[†]

*Instituto Superior Técnico, Departamento de Física, 1049-001 Lisboa, Portugal*

**Abstract**

In nearly all concepts invoked or proposed to change or shield gravity it is intuitively assumed that manipulation of gravity automatically leads to a breakthrough for propulsion. In this study it is shown, that even if gravity could be hypothetically controlled along the manipulation schemes outlined, the gains in terms of propulsion would be modest and lead to no breakthrough. Although the manipulation schemes presented are not exhaustive, they include the most straightforward ones from the current physics point of view.

[*] Principal Scientist, also Lecturer, Aerospace Engineering Department, A-1040, Vienna University of Technology, Phone: +43-50550-3142, Fax: +43-50550-3366, Email: martin.tajmar@arcs.ac.at, Member AIAA.
[†] Associate Professor, Phone: +351-21-841-7620, Fax: +351-21-841- 9118, Email: orfeu@cosmos.ist.utl.pt

## Nomenclature

| | |
|---|---|
| c | speed of light = $3 \times 10^8$ m.s$^{-1}$ |
| δ | inertial mass modification factor |
| ε | gravitational mass modification factor |
| F | force |
| G | gravitational constant = $6.67 \times 10^{-11}$ m$^3$.kg$^{-1}$.s$^{-2}$ |
| $g_0$ | standard gravitational acceleration = 9.81 m$^{-2}$ |
| γ | adiabatic index |
| $I_{sp}$ | specific impulse |
| $m_p$ | propellant mass |
| $\mu_p$ | molar propellant mass |
| R | universal gas constant = 8.31 J.mol$^{-1}$.K$^{-1}$ |
| $T_c$ | combustion chamber temperature |
| $v_p$ | propellant velocity |

# Introduction

As part of its advanced space propulsion program, NASA created the Breakthrough Propulsion Physics Program in 1996 to look for new concepts in propulsion that can ultimately enable interstellar travel[1,2]. Nearly all concepts proposed involve some method to alter gravity, such as warp drives[3], transient mass fluctuations[4,5], or gravitational shielding effects[6]. They all intuitively assume, without any justification, that manipulation of gravity would automatically lead to a breakthrough for propulsion.

Subsequentially, in 2001, the European Space Agency (ESA) funded a study to evaluate the concept of gravity control in light of current theories of gravity and field theory as well as to assess the scientific credibility of claims in the literature of anomalous gravitational experiments and phenomena[7]. Furthermore, the study was to analyse the impact on spacecraft propulsion of any degree of gravity control.

As anticipated, the first part of the study yielded no surprises as current experimental knowledge and bounds on the fundamental underlying principles of General Relativity and of the Standard Model of the Fundamental Interactions leave little room for the gravity control proposals that were analysed. Among these we could mention exotic concepts such as theories where gravitiy is due to interactions with the Zero-Point-Energy field[8,9], warp-drive mechanisms[3] or propulsion concepts based on Mach's principle[4,5]. None of the approaches examined proved fruitful. However, the second part of the study has turned out to be rather rich in new findings. Our main conclusion was that even if gravity could be controlled or modified, influence on spacecraft propulsion would be quite modest and would not lead to breakthroughs in the conceptual framework of presently known propulsion principles within the studied manipulation schemes. We regard this result of particular importance to the

Breakthrough Propulsion Physics community and should, in our opinion, be considered in any further work on the topic.

**Gravity Control**

Current knowledge of gravitational phenomena is accurately described by Einstein's Theory of General Relativity. The theory matches all known experimental data. Experimental windows of opportunity can be found in the untested ground of gravity-like forces below the millimetre scale or beyond $10^{12}$ km, or in violations of the Weak Equivalence Principle for anti-particles. Consistency between Quantum Mechanics and General Relativity that requires a still elusive quantum gravity theory, leads for instance to the quite novel framework of Superstring physics, and this theory may give rise to new effects, such as a deviation from the Weak Equivalence Principle at the $10^{-18}$ level[10]. The reader is referred to Ref. 7 for an extensive review of these issues (see also Ref. 11).

In addition to changes in the inertial or gravitational mass, there is one aspect of General Relativity that is more closely related to possible propulsion applications. In lowest order of the weak field limit, General Relativity can be restructured in a way that it closely resembles Maxwell´s equations[12]. In this context, there is a similarity between moving charges and moving masses, to which the classical laws of electromagnetism apply. For instance, a mass in motion generates a gravitational interaction that is similar to a magnetic field, the so-called gravitomagnetic field. This gravitomagnetic field can interact with other gravitomagnetic fields creating forces[13]. Such forces are very weak (usually $10^{-20}$ N or less in Earth like environments), but can in principle be detected by extremely sensitive gyroscopes such as the ones developed for NASA's Gravity Probe B (GP-B) mission. The frame dragging effect of the Earth was experimentally measured using the LAGEOS and LAGEOS II laser ranging satellites[14] and is currently being investigated by NASA's GP-B mission.

Induction laws such as the ones encountered in electromagnetism exist in this approximation of General Relativity making possible the conversion of gravitoelectric fields into gravitomagnetic fields and vice-versa. Effects related to the gravitomagnetic field are also referred to as frame-dragging or Lense-Thirring effect[12]. We shall analyse in more detail the effects for propulsion arising from such forces in a later section.

In quite general terms, we find that any scheme to hypothetically control gravity must fulfil at least one of the following conditions:

1. Existence of a new fundamental interaction of nature so to alter the effective strength of the gravitational coupling to matter. This implies violations of the Weak Equivalence Principle.

2. Existence of net forces due to the interplay between gravity and electrostatic forces in shielded experimental configurations, as found in the well-known Schiff-Barnhill effect[15].

3. Analogous effect for magnetic fields in quantum materials involving the gravitomagnetic field[16,17].

4. Physically altering the vacuum properties so to change the relative strength of known fundamental interactions of nature.

## Classical Spacecraft Propulsion and Definition of Terms

All classical propulsion systems rely on Newton's mechanics. Equations characterizing the performance of any propulsion system can be found in a large variety of textbooks such as in Ref. 18. In this section, we will recall basic equations that will be used for the analysis of influence on gravity control.

In general terms, assuming a constant propellant velocity, the force $F$ is defined as

$$\vec{F} = -|\dot{m}_p|\vec{v}_p, \tag{1}$$

where $\dot{m}_p$ is the propellant mass flow and $v_p$ the propellant velocity. To a good approximation, the specific impulse $I_{sp}$ is proportional to the propellant exhaust velocity. Neglecting the influence of the nozzle or the ambient atmosphere, the propellant velocity for a simple chemical thruster can be written as

$$v_p \cong \sqrt{\frac{2\gamma}{(\gamma-1)} \cdot \frac{RT_c}{\mu_p}}, \tag{2}$$

where $\gamma$ is the adiabatic index, $R$ is the universal gas constant, $T_c$ the temperature in the combustion chamber and $\mu_p$ the molar mass of the propellant. We clearly see, that the following scaling laws are valid:

$$F \propto m_p^{1/2}, \quad I_{sp} \propto v_p \propto m_p^{-1/2}. \tag{3}$$

Apart from electromagnetic thrusters, these scaling laws also apply for other propulsion systems such as electric thrusters (e.g. hall, ion, or field emission thrusters) to a good approximation. The amount of propellant needed for a given mission is derived from the trajectory analysis and is typically expressed as the required change of velocity from the spacecraft, $\Delta v$. The total requirement consists of several components,

$$\Delta v = \Delta v_g + \Delta v_{orbit} + \Delta v_{drag} - \Delta v_{initial}, \tag{4}$$

namely the $\Delta v_g$ to overcome the gravitational potential (e.g. from the Earth's surface to the required orbit), $\Delta v_{drag}$ due to drag from the atmosphere (usually around 0.1 km/s), $\Delta v_{orbit}$ giving the velocity increment to reach a certain orbit and $\Delta v_{initial}$ the initial velocity (e.g. due to the Earth's centrifugal force which on the Equator is about 0.4 km/s). The following equations are used to characterize the most dominant $\Delta v$ parts:

$$\Delta v_g = \sqrt{2GM\left(\frac{1}{r_{initial}} - \frac{1}{r_{final}}\right)}, \tag{5}$$

$$\Delta v_{orbit} = \sqrt{\frac{GM}{r}}, \tag{6}$$

where $G$ is the gravitational constant, $M$ the Earth's mass and $r$ the orbit's distance to the center of the Earth. For example, a Low-Earth Orbit (LEO) of 100 km altitude ($r_{initial}$ = 6400 km, $r_{orbit}$ = 6500 km), results in $\Delta v_{orbit}$ = 7.8 km/s and $\Delta v_g$ = 1.4 km/s, starting from Earth's surface. The total requirement would then be $\Delta v_{LEO}$ = (1.4+0.1+7.8-0.4) km/s = 8.9 km/s. The propellant mass necessary to meet the full $\Delta v$ requirement can be calculated using the well-known Tsiolkovski equation:

$$m_p = m_0 \cdot \left[1 - \exp\left(-\frac{\Delta v}{v_p}\right)\right] . \tag{7}$$

## Influence of Gravity Manipulation on Spacecraft Propulsion

In the following discussion we shall consider a series of hypothetical devices capable of manipulating mass (with influence in current propulsion systems) or generating an artificial gravitational field (new propulsion concept using gravity control). As already discussed, no such devices are known. Nevertheless, several combinations (during launch and in space) are going to be analysed and their influence on propulsion systems or as a force generator will be discussed. In order to avoid complexity (we are talking about hypothetical gravity control devices anyhow), we do not endow any special attributes to the gravity control manipulator and do not limit its manipulation to certain distances.

**Inertial Mass Modification**

Let us assume a device that can change the inertial mass of bodies in its interior, such as a launcher. If the spacecraft were to fire conventional chemical rockets or electric propulsion thrusters how would this device affect performance? We define an inertial mass modification factor $\delta$ and introduce it into the equations for the specific impulse and force for chemical thrusters:

$$I_{sp} \propto v_p \cong \sqrt{\frac{2\gamma}{(\gamma-1)} \cdot \frac{RT_c}{\mu_p \cdot \delta}} \propto \delta^{-1/2} \; , \tag{8}$$

and hence,

$$F = \delta m_p v_p \propto \delta^{1/2} \; . \tag{9}$$

Notice that these are the same scaling laws previously discussed. Thus, hypothetical inertial mass modification is similar to choosing propellant with a different molar mass. These laws are similar for all other classical propulsion systems. The only advantage would be that by manipulating inertial mass, the chamber temperature $T_c$ is not really changing. This might not be the case for choosing a lighter propellant in chemical thrusters (e.g. $H_2/O_2$ instead of Kerosen/$O_2$). If $T_c$ is lower, the specific impulse drops as well. Also the spacecraft mass can be reduced conventionally, e.g. by using lightweight structures. This will then reduce the Δv requirement accordingly – but not affect the propulsion system performance.

How is the *Δv* requirement influenced? The kinetic energy has now to be multiplied by the factor δ and one derives the modified equations:

$$\Delta v_g = \sqrt{2GM\left(\frac{1}{r_{initial}} - \frac{1}{r_{final}}\right) \cdot \frac{1}{\delta}} \propto \delta^{-1/2} \; , \tag{10}$$

and

$$\Delta v_{orbit} = \sqrt{\frac{GM}{r} \cdot \frac{1}{\delta}} \propto \delta^{-1/2} \; . \tag{11}$$

Hence, an inertial mass reduction would actually *increase* the $\Delta v$ requirement! For the extreme case of $\delta = 0$, the launcher would not move any more ($F = 0$) and accordingly the $\Delta v$ requirement would be infinite. Thus, for $0 < \delta < 1$, thrust $F$ would increase while the specific impulse $I_{sp}$ would decrease (see Fig. 1). This is similar to using a heavier propellant, e.g. solid boosters with alumna oxides instead of $H_2/O_2$ chemical thrusters.

An interesting result is that Tsiolkovski's Eq. (7) is not affected at all by a modification on the inertial mass. Indeed, to a good approximation one can neglect $\Delta v_{drag}$ and $\Delta v_{initial}$ so that the total $\Delta v \propto \delta^{1/2}$. Furthermore, one can see that all $\delta$ factors are cancelled out in Eq. (7). Therefore, whatever is the inertial mass modification, the amount of propellant will not be changed, or in other words, the change in the propulsion performance is always counterbalanced by the change in the $\Delta v$ requirement!

We conclude that a modification on the inertial mass is of *no* of interest for propulsion.

**Gravitational Mass Modification**

We turn now to the analysis on the modification on the gravitational mass. Such modification does not influence propulsion performance (thrust, specific impulse), only parts of the $\Delta v$ requirement are affected in this case. By using the gravitational mass modification factor dividing the gravitational by the intertial mass, $\varepsilon = m_g/m_i$, we get

$$\Delta v_g = \sqrt{2GM\varepsilon \left( \frac{1}{r_{initial}} - \frac{1}{r_{final}} \right)} \propto \varepsilon^{1/2} , \qquad (12)$$

and

$$\Delta v_{orbit} = \sqrt{\frac{GM}{r} \cdot \varepsilon} \propto \varepsilon^{1/2} . \tag{13}$$

Neglecting as before $\Delta v_{drag}$ and $\Delta v_{initial}$, we obtain, $\Delta v \propto \varepsilon^{1/2}$, approximately. The modified Tsiokovski's equation can then be written as

$$m_p = m_0 \cdot \left[1 - \exp\left(-\frac{\Delta v \cdot \varepsilon^{1/2}}{v_p}\right)\right] . \tag{14}$$

If the gravitational mass would be reduced, that is $\varepsilon < 1$, the $\Delta v$ requirement would drop and less propellant mass would be required. For a satellite orbiting the Earth, our modified definition of $\Delta v_{orbit}$ in Eq. (13) also increases the time it needs to make a full orbit (less speed to cycle the Earth). This may not be important for launching interplanetary probes, but affects for instance, remote sensing or Geostationary Orbit (GEO) satellites which require cycling Earth with a certain speed for the purpose of mapping and telecommunication. In such cases the full $\Delta v_{orbit}$ ($\varepsilon = 1$) would have to be applied with no reduction, which then in turn causes an unstable orbit since the centrifugal force would be higher than the gravitational pull. To counterbalance this effect, two different strategies are possible:

- Launch of the spacecraft with the gravitational mass manipulator (GMM) activated. Once in orbit, deactivate GMM and apply full $\Delta v_{orbit}$ using a propulsion system. The centrifugal force is then balanced by the gravitational push.

- GMM is always active and use propulsion system to counterbalance the higher centrifugal force to stay in orbit.

As an example we consider a scientific satellite for a planetary target ($\Delta v$ = 10 km/s) with a chemical thruster ($v_p$ = 3500 m/s). In this case we do not need a full $\Delta v_{orbit}$ and our above mentioned concerns do not apply. Fig. 2 the $\Delta v$ and propellant mass ratio ($m_p/m_0$) reduction are plotted as a function of $\varepsilon$. For $\varepsilon$ = 0.05, only half of the propellant would be required. Of course, shielding very close to $\varepsilon$ = 0 would reduce the propellant consumption to zero and a very direct trajectory to the target would be possible. Applying Newton's laws, the trip time to reach a required $\Delta v$ scales with $\varepsilon^{1/2}$. Therefore, shielding of gravitational mass does reduce the trip time, however, even shielding of $\varepsilon$ = 0.05 only reduces trip time by a factor of 4.5. In conclusion, unless almost total shielding can be achieved, there is no breakthrough in the overall trip time.

Let us assume the extreme case where the gravitational mass vanishes and thus $\Delta v_g$ vanishes. The $\Delta v$ requirement, still including the full $\Delta v_{orbit}$ for typical LEO satellites, would then change from Eq. (4) to

$$\Delta v = \Delta v_{orbit} + \Delta v_{drag} - \Delta v_{initial} , \qquad (15)$$

Considering again our previous 100 km LEO example, the case of $\Delta v_g$ = 0 would reduce the total $\Delta v$ requirement from the initial 8.9 km/s by 1.4 km/s, that is down to 7.5 km/s. So a launcher, although with less propellant, is still required. The higher the original $\Delta v_g$, the higher the possible reduction of the total $\Delta v$. GEO satellites with a high altitude of 42,160 km could reduce in this case from a total $\Delta v$ of about 13 km/s to 3 km/s which would require a much smaller propulsion system reducing launch costs drastically. Only in the case where $\Delta v_{orbit}$ is not important, as in e.g. interplanetary spacecraft, we can assume a full reduction to

$$\Delta v = \Delta v_{drag} - \Delta v_{initial}, \tag{16}$$

If the spacecraft is close to the Equator, then $\Delta v_{initial} \approx 0.4$ km/s from the spinning Earth is higher than $\Delta v_{drag} \approx 0.1$ km/s and so it would start lifting by itself. That could lead to completely new launch strategies and would certainly be a breakthrough.

How far are we away from such a possible breakthrough? The experimental verification of the Weak Equivalence Principle indicates that $\varepsilon = 1 \pm 5 \times 10^{-13}$. On the other hand, string theory predicts[10] that $\varepsilon = 1 \pm 1 \times 10^{-18}$. Obviously, these values do not influence propulsion at all. As previously outlined, significant deviations from $\varepsilon = 1$ are not forseen, unless, for instance, the behaviour of antiparticles on a gravitational field is substantially different from the one of particles.

Even if situations where $\varepsilon \neq 1$ can be engineered, they would have to compete with concepts such as electric propulsion, which can lower the propellant consumption already by 90% with much higher propellant velocities. There are also technologies and concepts available which can reduce the $\Delta v$ requirement and trip time. One example is to reduce the gravitational potential $\Delta v_g$ simply by launching from a mountain or to supply a higher initial velocity $\Delta v_{initial}$ by launching from an airplane (e.g. the Pegasus rocket from Orbital Sciences Corporation is currently being launched from an aircraft for reduction of the $\Delta v$ requirement). NASA, for instance, is developing an initial acceleration rail for future spacecraft that uses superconducting magnetic levitation (MAGLIFTER) to fire the rocket engine after the spacecraft reached a speed of 300 m/s. For an extreme summary of launch assist technology, see Ref. 19. An extreme concept was out forward by Arthur C. Clarke[20], who proposed to built an ultra-high tower, which was named space elevator, for lifting spacecraft to orbits as high as 100 km.

**Dipolar Gravito-Electric Field**

The possibility to generate artificial gravitational fields would enable new concepts for space propulsion. According to the induction laws of gravitomagnetism[12,13], such an artificial field would look like a dipole composed by gravito-electric fields. The most straightforward propulsion concept would be to interact with the Earth's gravitational field, as outlined in Fig. 3. Such a field configuration could be used to generate a torque orientating the spacecraft parallel to the planet gravitational field. In analogy with electromagnetism, we can write

$$\vec{\tau}_g = \vec{\mu}_g \times \vec{g} , \qquad (17)$$

where $\tau_g$ is the torque, $\mu_g$ the gravitoelectric dipole momentum and $g$ the Earth's gravitational field. Since much simpler conventional concepts are available, we shall not elaborate further on this idea. The simplest one is the so-called gravity boom[21], which is used in almost all small satellites. It consists of a test mass along a deployable boom attached to the spacecraft. Due to the gravitational gradient, the boom always points the satellite towards Earth. In addition to its simplicity, such a device does not require any power at all.

We want to stress that such a dipolar gravito-electric field is not at all a gravitational dipole, which is a mathematical object composed of a positive and negative mass. A gravitational dipole would be self-accelerating[22], however, it was been clearly shown that negative masses are forbidden in gravitational physics and that a propulsion system based on such a concept does not make sense[7]. The impossibility of negative mass in our universe has been shown by the positive energy theorem[23,24], which is known to apply to all known matter in both normal and extreme cosmological situations. Violations of the energy conditions are possible in quantum theories, but the effects are fleeting and require exotic equations of state[25-27].

**Wire-like Gravitomagnetic Field**

Through the use of a "wire-like" gravitoelectric field or a gravitomagnetic field, one could generate a gravitational analogue of the Lorentz force. This could be used as a propulsion system (see Fig. 4) interacting with a planet's gravitomagnetic field. By using Gravitoelectromagnetism, one can conceive a gravitational analogue of an electrodynamic tether. The force $F$ produced by a wire in a gravitational field can be expressed as

$$F = \frac{B_g}{c} \cdot I_m \cdot l , \qquad (18)$$

where $B_g$ is the gravitomagnetic field (for Earth $B_g = 3.3 \times 10^{-14}$ rad/s), $I_m = dm/dt$, is the mass current and $l$ the length of the wire. For $I_m = 3000$ kg/s (for example by pumping liquid metal through a tube), a length of 1 km and the Earth's gravitomagnetic field, this force is equal to 0.1 µN. Comparing this unrealistic concept (huge $I_m$) with an electrodynamical tether used for satellites, with a typical power consumption of only 2.4 kW, a force of 0.36 N is produced in LEO. Therefore, a gravitational analogue adds no extra benefit to current tether technology.

Even if much larger gravitomagnetic fields than the one produced by the Earth could be produced, e.g. using superconductors as proposed in Ref. 17, this conclusion does not change.

# Conclusion

We can summarize our results as follows. Within the context of propulsion devices based on the reaction principle, our study reveals that control of gravity, even if achievable, would not imply in a breakthrough for propulsion, even though it could be of major importance for e.g. possible microgravity applications on Earth. This is of course only valid for the manipulation schemes outlined. Although the manipulation schemes presented are not exhaustive, they include the most straightforward ones from the current physics point of view.

More concretely our analysis reveals that modification of inertial mass would bring no influence at all, and the modification of gravitational mass would have to compete with classical-launch assist technologies such as launching from an airplane, top of a mountain, or in an extreme case, from an ultra-high tower. Moreover, the use of gravitomagnetic or gravitoelectric fields for propulsion would not bring any extra benefit when compared to classical electrodynamical tethers or gravity booms. Our conclusions are summarized in Table 1. We believe that the result of our analysis is a valuable input to ongoing breakthrough propulsion activities in the Unites States and Europe.

# Acknowledgement

This work was carried out under ESTEC Contract 15464/01/NL/Sfe, funded by ESA General Studies Programme. We would like to thank Clovis De Matos and J.C. Grenouilleau for varios important dicussions and for their continuous support throughout the study.

# References

[1]Millis, M. "Challenge to Create the Space Drive," Journal of Propulsion and Power, Vol. 13, No. 5, 1997, pp. 577-682

[2]Millis, M., "NASA Breakthrough Propulsion Physics Program," Acta Astronautica, Vol. 44, Nos. 2-4, 1999, pp. 175-182

[3]Alcubierre, M., "The Warp Drive: Hyper-Fast Travel with General Relativity," Classical and Quantum Gravity, Vol. 11, 1994, pp. L73-77

[4]Woodward, J.F., Mahood, T., and March, P., "Rapid Spacetime Transport and Machian Mass Fluctuations: Theory and Experiment," AIAA Paper 2001-3907, 2001

[5]Woodward, J.F., and Mahood, T., "Gravity, Inertia, and Quantum Vacuum Zero Point Energy Fields," Foundations of Physics, Vol. 31, No. 5, pp. 819-835

[6]Robertson, G.A., Litchford, R., Thompson, B., and Peters, R., "Exploration of Anomalous Gravity Effects by Magnetized High-Tc Superconducting Oxides," AIAA Paper 2001-3364, 2001

[7]Bertolami, O., and Tajmar, M., "Gravity Control and Possible Influence on Spacecraft Propulsion: A Scientific Study," ESTEC Contract Report 15464/01/NL/Sfe, Noorwijk, 2002

[8]Haisch, B., Rueda, A., and Puthoff, H.E., "Inertia as a zero point field Lorentz force," Physical Review A, Vol. 49, No. 2, 1994, pp. 679-694

[9]Dobyns, Y., Rueda, A., and Haisch, B., "The Case for Inertia as a Vacuum Effect: A Reply to Woodward and Mahood," Foundation of Physics, Vol. 30, 2000, pp. 59-80

[10]Damour, T., Polyakov, A.M., "String Theory and Gravity," General Relativity and Gravitation, No. 26, 1994, pp. 1171

[11]Bertolami, O., de Matos, C.J., Grenouilleau, J.C., Minster, O., and Volonte S., "Perspectives in Fundamental Physics in Space," Los Alamos Physics Archive gr-qc/0405042, 2004

[12]Forward, R.L., "General Relativity for the Experimentalists," Proceedings of the IRE, Vol. 49, 1961, pp. 892-904

[13]Forward, R.L., "Guidelines to Antigravity," American Journal of Physics, Vol. 31, 1963, pp. 166-170

[14]Ciufolini I., Pavlis E., Chieppa F., Fernandes-Vieira E., Perez-Mercader J., "Test of General Relativity and Measurement of the Lense-Thirring Effect with two Earth Satellites," Science, Vol. 279, 1998, pp. 2100-2103

[15]Schiff, L.I., Barnhill, M.V., "Gravitation-Induced Electric Field Near a Metal," Physical Review, Vol. 151, No. 4, 1966, pp. 1067-1071

[16] DeWitt, B.S., "Superconductors and Gravitational Drag," Physical Review Letters, Vol. 16, No. 24, 1966, pp. 1092-1093

[17] Tajmar, M., and de Matos, C.J., "Gravitomagnetic Field of a Rotating Superconductor and of a Rotating Superfluid," *Physica C*, Vol. 385, No. 4, 2003, pp. 551-554

[18] Sutton, G.P., and Biblarz, O., "Rocket Propulsion Elements," 7th Edition, John Wiley & Sons, 2001

[19] Tajmar, M., "Advanced Space Propulsion Systems," Springer, Wien-NewYork, 2002

[20] Clarke, A.C., "The Space Elevator: Thought Experiment or Key to the Universe?," Earth Oriented Application of Space Technology, Volume I, 1981, pp. 39-48

[21] Larson, W.J., and Wertz, J.R., "Space Mission Analysis and Design," Kluwer Academic Publisher, 1991, pp. 345

[22] Forward, R.L., "Negative Matter Propulsion," Journal of Propulsion and Power, Vol. 6, No. 1, 1990, pp.28-37

[23] Witten, E., "A New Proof of the Positive Energy Theorem," Commun. Math. Phys., Vol. 80, 1981, pp. 381-392

[24] Schoen, P., and Yau, S.T, "Positivity of the Total Mass of a General Space-Time," Physical Review Letters, Vol. 43, 1979, pp. 1457-1459

[25] Morris, M.S., Thorne, K.S., and Yurtsever, U., "Wormholes, Time Machines, and the Weak Energy Condition," Physical Review Letters, Vol. 61, 1988, pp. 1446-1449

[26] Deser, S., and Jackiw, R., and 't Hooft, G., "Physical cosmic strings do not generate closed timelike curves," Physical Review Letters, Vol. 68, 1992, pp. 267-269

[27] Cho, Y. M. and Park, D. H., "Closed time-like curves and weak energy condition," Physics Letters B, Vol. 402, No. 1-2, 1997, pp. 18-24

|  | F | $I_{sp}$ | $\Delta v$ | $m_p/m_0$ | **Compare with Present Technology** |
|---|---|---|---|---|---|
| Inertial Mass Modification (by Factor $\delta$) | $\delta^{1/2}$ | $\delta^{-1/2}$ | $\delta^{-1/2}$ | - | Has no influence at all |
| Gravitational Mass Modification (by Factor $\varepsilon$) | - | - | $\varepsilon^{1/2}$ | $1 - \exp\left(-\dfrac{\Delta v \varepsilon^{1/2}}{v_p}\right)$ | • Requires very high shielding $\varepsilon \approx 0$ <br> • Electric Propulsion already saves up to 90% of propellant <br> • Launch from ultra-high tower or aircraft can save $\Delta v$ too |
| Dipolar Gravito-electric / magnetic Field | Only Torque | - | - | - | Gravity-booms are simple and require no power at all |
| Wire-like Gravitomagnetic Field | $< 10^{-7}$ N | - | - | - | Electrodynamical tethers have much higher thrusts |

**Table 1** Summary of Influence on Propulsion Systems for Gravity Control and Gravitoelectromagnetic Propulsion System

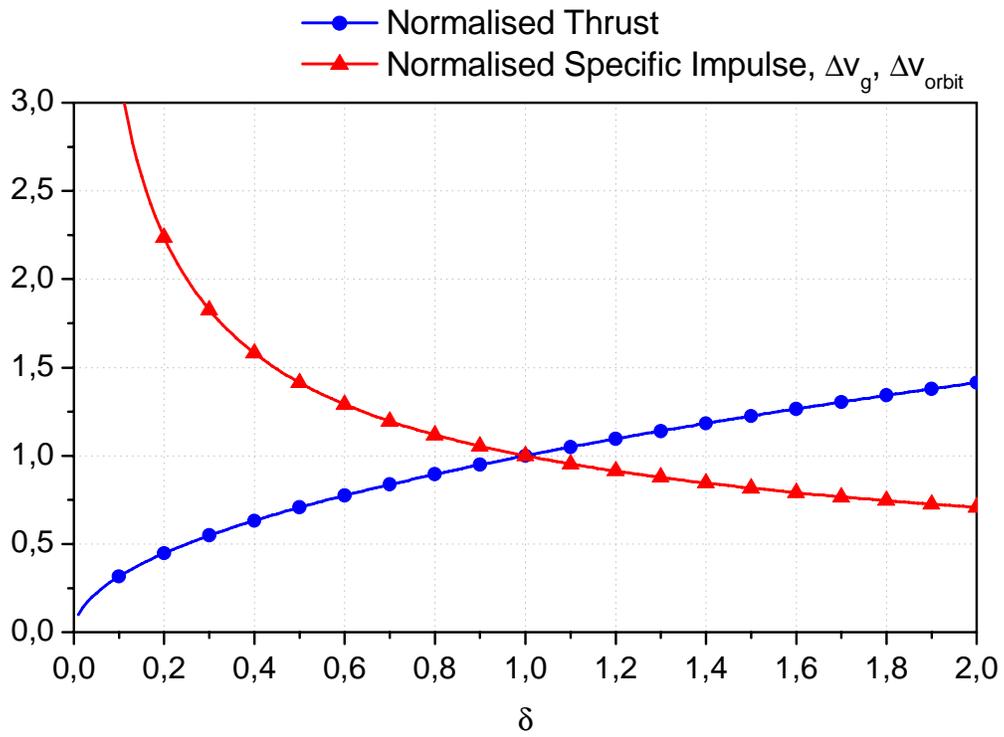

**Figure 1**  Influence of Inertial Mass Modification Factor δ on Propulsion Parameters

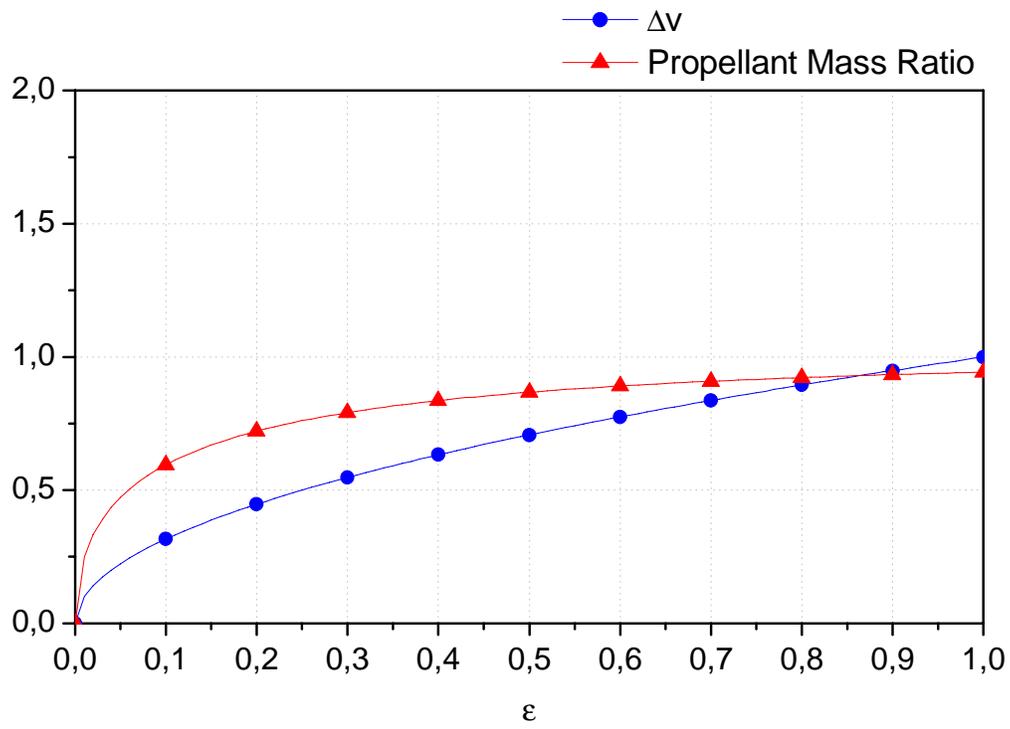

**Figure 2**  Influence of Gravitational Mass Modification Factor ε on Propulsion Parameters

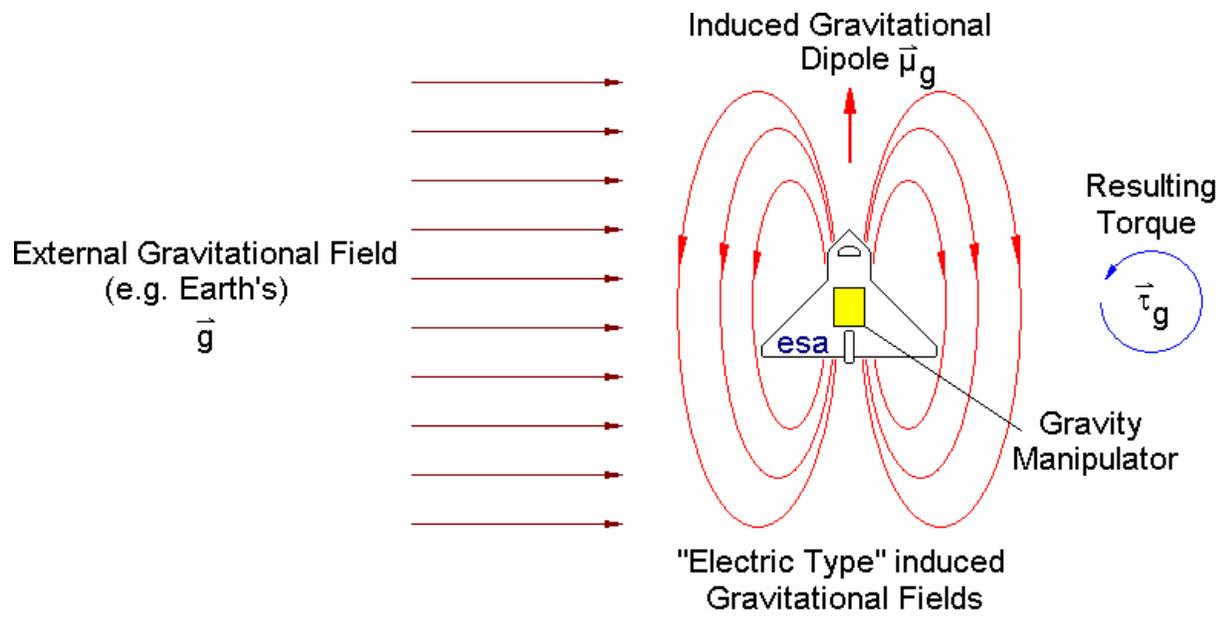

**Figure 3**  Illustration of Dipolar Gravito-Electric / Magnetic Field Propulsion

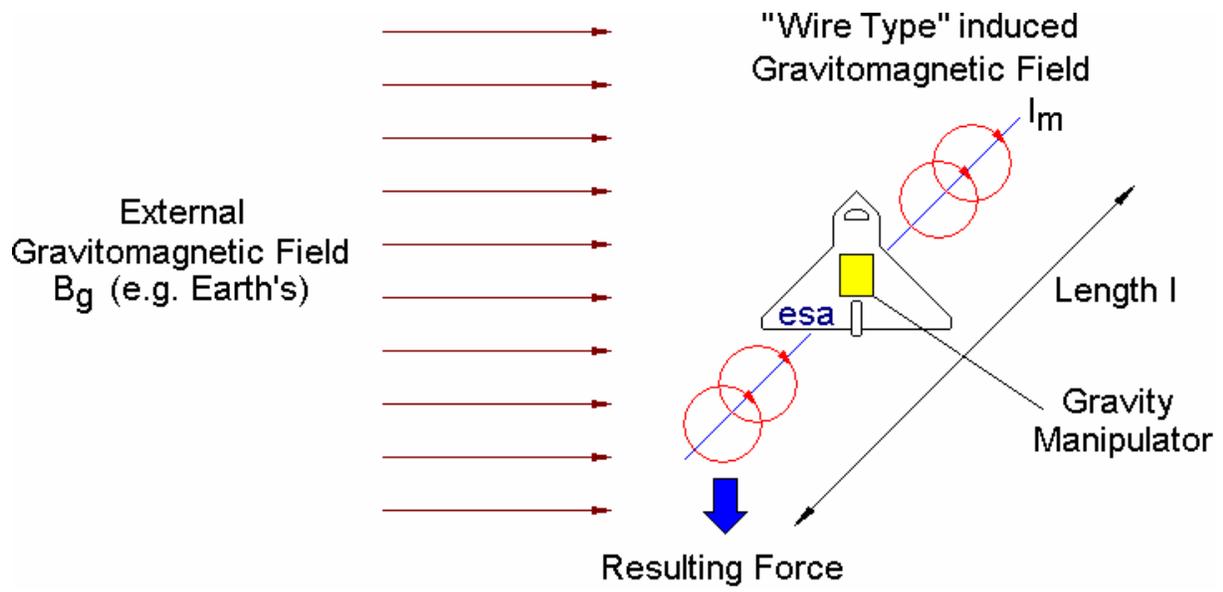

**Figure 4**  Illustration of Wire-like Gravitomagnetic Field Propulsion